\documentclass[12pt]{article}

\usepackage{setspace}
\usepackage[paper=letterpaper]{geometry}
\usepackage{graphicx}
\usepackage{authblk}
\usepackage{amsmath} 
\usepackage{amssymb}  
\usepackage{xcolor}
\usepackage{algorithmic}
\usepackage{chemmacros}
\usepackage{algorithm}
\usepackage{scalerel}
\usepackage{chemmacros}
\usepackage{array,multirow}
\usepackage{tabularx}
\usepackage{etoolbox}
\usepackage{array}
\newcolumntype{L}[1]{>{\centering\let\newline\\\arraybackslash\hspace{0pt}}m{#1}}

\usepackage{breqn}

\def\msquare{\mathord{\scalerel*{\Box}{gX}}}
\makeatletter
\def\endfigure{\end@float}
\def\endtable{\end@float}
\makeatother

\usepackage[caption=false]{subfig}
\usepackage{etoolbox}

\AtBeginEnvironment{figure}{\setlength{\captionwidth}{0.8\linewidth}}
\AtBeginEnvironment{table}{\setlength{\captionwidth}{0.8\linewidth}}
\renewcommand\appendix{\par
  \setcounter{section}{0}
  \setcounter{subsection}{0}
  \setcounter{figure}{0}
  \setcounter{table}{0}
  \renewcommand\thesection{Appendix \Alph{section}}
  \renewcommand\thefigure{\Alph{section}\arabic{figure}}
  \renewcommand\thetable{\Alph{section}\arabic{table}}
}
    
\begin{document}

\title{A Kalman Filter Approach for Biomolecular Systems with Noise Covariance Updating}
\author[1]{Abhishek Dey}
\author[2]{Kushal Chakrabarti}
\author[3]{Krishan Kumar Gola}
\author[1]{Shaunak Sen}
\affil[1]{Department of Electrical Engineering,\\
        Indian Institute of Technology Delhi, Hauz Khas, New Delhi 110016, INDIA.}
\affil[2]{Department of Electrical and Computer Engineering,\\ 
        University of Maryland, College Park, MD - 20742, USA.}
\affil[3]{Department of Electrical and Electronics Engineering,\\ 
        Meerut Institute of Engineering and Technology, Meerut - 250005, India.}
\maketitle
\noindent
\textbf{Abstract.} 
An important part of system modeling is determining parameter values, particularly for biomolecular systems, where direct measurements of individual parameters are typically hard. While Extended Kalman Filters have been used for this purpose, the choice of the process noise covariance is generally unclear. In this chapter, we address this issue for biomolecular systems using a combination of Monte Carlo simulations and experimental data, exploiting the dependence of the process noise covariance on the states and parameters, as given in the Langevin framework. We adapt a Hybrid Extended Kalman Filtering technique by updating the process noise covariance at each time step based on estimates. We compare the performance of this framework with different fixed values of process noise covariance in biomolecular system models, including an oscillator model, as well as in experimentally measured data for a negative transcriptional feedback circuit. We find that the Extended Kalman Filter with such process noise covariance update is closer to the optimality condition in the sense that the innovation sequence becomes white and in achieving balance between the mean square estimation error and parameter convergence time. The results of this chapter may help in the use of Extended Kalman Filters for systems where process noise covariance depends on states and/or parameters.

\section{INTRODUCTION}
\label{sec1}

Differential equation-based models require information about system parameters. This can be obtained through first
principles, direct measurement, or indirect parameter estimation. Biomolecular system models can be represented to different levels of approximation starting from the Chemical Master Equation, including the Langevin~\cite{khanin2008chemical} or Fokker-Planck approximations~\cite{van1992stochastic}, both stochastic, and the mass-action based deterministic approximations~\cite{del2015biomolecular}. These models are in general nonlinear and noisy~\cite{mcadams1999sa}, and may have many parameters that are typically hard to measure directly. Moreover, the noise associated with these models may itself depend upon unknown parameters that makes the parameter estimation for biomolecular systems challenging. Therefore, there is a need to adapt existing estimation methods when applied to these systems.\\
\indent Many parameter estimation techniques have been used in biological contexts, including search and global optimization techniques \cite{ashyraliyev2008parameter, zhan2011parameter}. Techniques based on state observers or Kalman filters~\cite{kalman1960new} have been used extensively as well in these contexts~\cite{murty2001useI, murty2001useII}. In general for these nonlinear systems
extended~\cite{dochain2003state, sun2008extended, wang2009extended} or unscented~\cite{meskin2013parameter, mansouri2014state} versions of Kalman filter formulation have been used. There is a rich background of using Extended Kalman Filters (EKF) to estimate state and parameters in \textit{Escherichia coli} fed-batch cultures~\cite{dubach1992application, veloso2009monitoring, dewasme2013experimental}. The Extended Kalman Filter has also been used to estimate moments of reacting species for biomolecular systems in a stochastic framework~\cite{ruess2011moment}. 
For these systems, the experimental measurements are in discrete time intervals and the models may be in continuous time. As such, these belong to the class of continuous-discrete systems for which EKF formulations for additive and multiplicative noise have been developed~\cite{jazwinski1970stochastic, jimenez2002linear}.
A particular filter of this continuous-discrete class, called the Hybrid Extended Kalman Filter (HEKF), has been used for parameter estimation in heat shock response in \textit{E. coli} and in repressilator using sampled data~\cite{lillacci2010parameter}. 
Typically, a fixed structure of the process noise covariance matrix is used in these contexts, chosen based on experience or trial and error.
In general, it has been noted, the process noise covariance matrix is an important design parameter for Kalman filter, but hard to tune due to unknown noise statistics~\cite{speyer2008stochastic, anderson1979optimal}.\\
\indent There are at least three striking aspects in estimating parameters in biomolecular systems using Kalman filters. One is that these systems are typically nonlinear, so EKF formulations may be needed. Two, these systems are inherently noisy which makes estimation of parameters challenging. Three, the process noise may depend on state and parameters which may be unknown. Whether the dependence of the process noise on parameters and states can be used to benefit the estimation process is generally unclear.\\
\indent Here, we investigate how the information about the dependence of process noise on parameters and states can be used in the estimation process.
To address this, we adapt the HEKF technique and apply it to estimate parameters from Monte Carlo simulation data of simple models of biomolecular circuits, including those with a limit cycle solution, as well as from experimental data of a negative transcriptional feedback circuit.
We emphasize the choice of the process noise covariance by updating, based on the Chemical Langevin Equation, the process noise covariance at each time step to reflect the updated state and parameters.
We find that the innovation sequence with the process noise covariance update can be white, in contrast to the case with a fixed covariance, indicating that the filter is closer to the optimality condition.
Further, this updated process noise covariance naturally provides a balance between mean squared estimation error and parameter convergence time compared to a fixed one.
These results can help in implementing the Kalman filter for biomolecular systems as well as for other systems where process noise is known to depend on states and parameters.

\section{MATHEMATICAL FRAMEWORK}
\label{sec2}
\subsection{Hybrid Extended Kalman Filter}

The experimental data for biomolecular circuits is typically acquired at discrete, uniform time intervals,
\begin{equation}
y_{t_k} = x_{t_k} + v_{t_k},
\label{measurement}
\end{equation}
where $y$ is the output vector, $x$ is the state vector, $v$ is the measurement noise, and the subscript $t_k$ denotes uniformly spaced discrete time instants. A mathematical model of such a biomolecular circuit may take multiple forms. A widely used mathematical modelling framework is the ordinary differential equation-based mass action kinetics,
\begin{equation}
\frac{dx}{dt} = f(x,\theta),
\label{ODE}
\end{equation}
where $[ f(x,\theta)]_i = \sum_{j=1}^{m} \zeta_{ij}a_j(x,\theta), i = 1,2,..,n$, the vector $x \in \mathcal{R}^n$ represents the biomolecular concentrations, $\theta \in \mathcal{R}^p$ represent parameters such as rate constants of biomolecular reactions, $a_j$ are the propensity functions, $\zeta_{ij}$ are the elements of the state change vector coming from the stoichiometry matrix, and $j = 1,2, .., m$ are the reaction channels~\cite{del2015biomolecular}. 
Another, more first principles-based method is to formulate a Chemical Master Equation where the variables are the probability distribution functions for various biomolecular concentrations at a time instant~\cite{van1992stochastic}. 
This latter formulation may be exactly simulated using Gillespie's Algorithm to obtain time trajectories of the biomolecular reactions~\cite{gillespie1977exact}. 
The former is a purely deterministic approach, whereas the latter includes the inherent stochasticity due to process noise. 
An intermediary approach, combining these two aspects is the Chemical Langevin Equation~\cite{del2015biomolecular, gillespie2000chemical}, which is an It\^{o} stochastic differential equation,
\begin{equation}
dx = f(x, \theta)dt + g(x, \theta)d\beta_t,
\label{SDE}
\end{equation}
where $[g(x, \theta)]_{ij} = \zeta_{ij}\sqrt{a_{ij}(x,\theta)}$, and $\{\beta_t\}$ is an n-vector Brownian motion process with $\mathbb{E}\{d\beta_t d\beta_t^T\} = Q(t)dt, Q(t) = I_{n \times n}$. 
The deterministic part in Eqn. (\ref{SDE}) matches that in Eqn. (\ref{ODE}). The stochastic part is different and gives the process noise. We note that the stochastic part may depend on the states and parameters.

The experimental data in combination with a model may be used to estimate unknown parameters. A standard way to do this is to use an EKF. In particular, the Kalman filter can be extended by redefining the parameter as a state and then by linearizing around a suitable estimate. 
The redefinition of the unknown parameters (denoted $\theta^u \in \mathcal{R}^s$) as states gives,
\begin{align}
& dx^E = f^E(x^E,\theta^v)dt + g^E(x^E,\theta^v)d\beta_t^E, \nonumber \\
& y_{t_k} = Mx^E_{t_k} + v_{t_k},
\label{extended system}
\end{align}
where $x^E = (x, \theta^u)$ is the extended state, $\theta^v \in \mathcal{R}^r, r = p-s$ are the known parameters, $f^E(x^E,\theta^v) = [f(x,\theta);0]$, $g^E(x^E,\theta^v) = \left[
\begin{array}{c|c}
g(x,\theta) & 0 \\
\hline
0 & \eta
\end{array}
\right]$, $\eta \in \mathcal{R}^{s \times s}$ is a block diagonal matrix with small positive diagonal entries, $\{\beta_t^E \}$ is an $(n+s)$-vector Brownian motion process with $\mathbb{E}\{d\beta_t^E {d\beta_t^E}^T\} = Q^E(t)dt, Q^E(t) = I_{(n+s) \times (n+s)}$, $y_{t_k} \in \mathcal{R}^z$ is the measurement vector at $t_k$ and $M \in \mathcal{R}^{z \times n}$. 
This is then linearized around an estimate,
\begin{align}
& d\delta x^E = F(t)\delta x^E dt + G(t)d\beta_t, \nonumber \\
& \delta y_{t_k} = M\delta x^E_{t_k} + v_{t_k}.
\label{CDsystem}
\end{align}
For such a continuous-discrete system, the Kalman-Bucy filter is given as follows~\cite{jazwinski1970stochastic}. Between observations, the estimates satisfy the differential equations,
\begin{align}
& \frac{d\hat{x}_t^{t}}{dt} = F(t)\hat{x}_t^t, \nonumber \\
& \frac{dP_t^t}{dt} = F(t)P_t^t + P_t^tF^T(t) + G(t)Q(t)G^T(t),
\label{predict}
\end{align}
for $t_{k-1} \leq t < t_k$. At an observation at $t_k$, they satisfy the difference equations,
\begin{align}
& \hat{x}_{t_k}^{t_k^+} = \hat{x}_{t_k}^{t_k^-} + K_{t_k}(y_{t_k} - M_{t_k}\hat{x}_{t_k}^{t_k^-}), \nonumber \\
& P_{t_k}^{t_k^+} = [ I - K_{t_k}M_{t_k} ]P_{t_k}^{t_k^-}[ I - K_{t_k}M_{t_k}]^T + K_{t_k}RK^T_{t_k},
\label{correct}
\end{align}
where the Kalman gain is given by,
\begin{equation}
K_{t_k} = P_{t_k}^{t_k^-}M^T_{t_k}[ M_{t_k}P_{t_k}^{t_k^-}M^T_{t_k} + R]^{-1}.
\label{Kalman gain}
\end{equation}
Typically, the prediction in Eqn. (\ref{predict}) is done by using the original equation in Eqn. (\ref{extended system}) i.e. $dx^E = f^E(x^E,\theta^v)dt$.

\begin{figure}[thpb]
      \centering
   \hspace{-0.1cm} \includegraphics[scale=0.41]{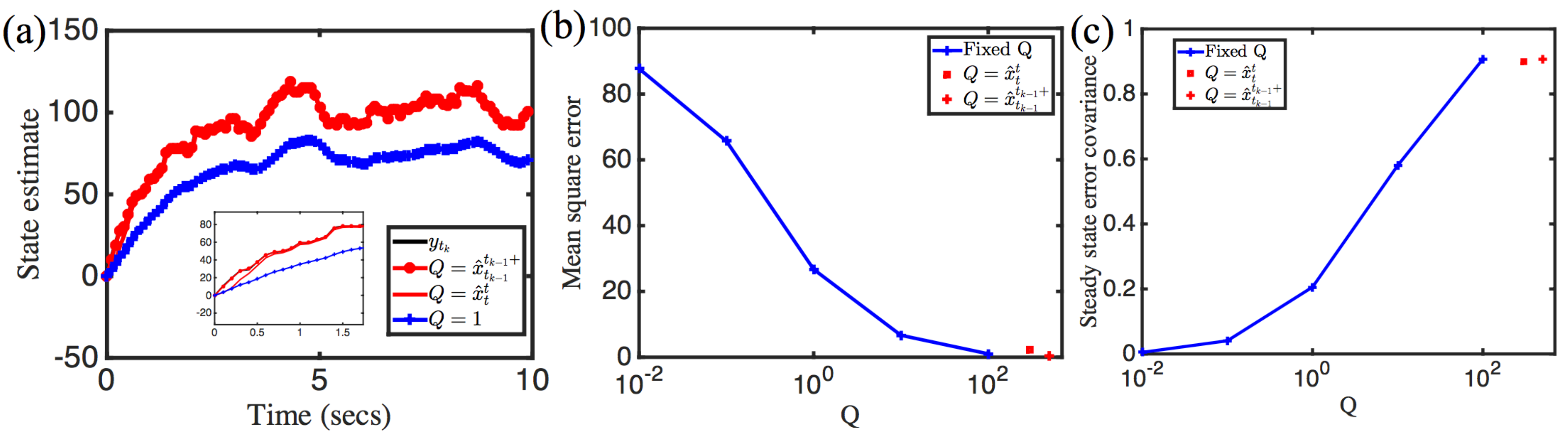}
      \caption{State estimation for Example 1. Comparison of a. state estimates (inset shows zoomed in version), b. mean square estimation error for the three cases of process noise covariance, c. corresponding error covariances.}
      \label{fig:illus_kf}
   \end{figure}

The choice of the process noise covariance is generally unknown and is understood to be challenging and crucial for filter performance. The standard approach is to use a constant covariance matrix for the process noise. Here, we investigate the consequence of using an effectively time-dependent noise covariance obtained by approximating the process noise by its value at the estimate using the information given in the Langevin Equation (Eqn. (\ref{SDE})). We note that there are two ways of doing this. One way, as per the linearization, is to replace process noise with its value at the best previous estimate, 
\begin{equation}
Q_{t_k} =  G(t_k)G^T(t_k) = \left[
\begin{array}{c|c}
Q_{i,k} & 0 \\
\hline
0 & \eta
\end{array}
\right],
\label{updatedQ}
\end{equation}
where $Q_{i,k} = \sum \zeta_{ij}^2a_j(\hat{x}_{t_{k-1}}^{t_{k-1}^+}), i = 1,2,..,n$ and $\eta \in \mathcal{R}^{s \times s}$ are block diagonal matrices. Another way is to replace process noise with its value at the estimate $\hat{x}^t_t$, in line with using the complete state equation in Eqn. (\ref{extended system}) in place of the state esimate equation in Eqn. (\ref{predict}). This framework illustrated in Algorithm \ref{algo} may be more appropriate for biomolecular systems and avoid the need to make an ad hoc guess.
\begin{algorithm}[H]
\begin{algorithmic}[1]
\caption{Algorithm for HEKF with process noise covariance update}\label{algo}
\STATE \textit{Initialisation} : $\hat{x}_0^{+}, P^+_0, R$
\FOR {$k = 1$ to $N$ (number of data points)}
\STATE \textbf{Update process noise covariance} 
\STATE Predict estimates, $\hat{x}_t^{t}, P_t^t$ (Eqn. (\ref{predict}))
\STATE Compute Kalman gain (Eqn. (\ref{Kalman gain}))
\STATE Correct estimates, $\hat{x}_{t_k}^{t_k^+}, P_{t_k}^{t_k^+}$ (Eqn. (\ref{correct}))
\ENDFOR
\end{algorithmic}
\end{algorithm}
We use an example of state estimation to illustrate this scheme as well as this difference.\\
\textit{Example 1}. Consider a scalar linear It\^{o} stochastic differential equation,
\begin{align}
& dx = (\alpha - x)dt - \sqrt{x}d\beta_t, \nonumber \\
& y_{t_k} = x_{t_k} + v_{t_k}.
\label{linear example}
\end{align}
It is desired to estimate the state using output measurements. To generate the data, this is simulated using the Euler-Maruyama method with a time step of $0.001$ for $10000$ time instants. 
Every $100$th point is taken and a measurement noise corresponding to a zero mean unit variance white Gaussian noise is added to it to obtain the data. Using Eqn. (\ref{predict}), the Kalman-Bucy filter equations between the observations are,  
\begin{align}
& \frac{d\hat{x}_t^{t}}{dt} = -\hat{x}_t^{t}, \nonumber \\
& \frac{dP_t^t}{dt} = -2P_t^t + Q.
\end{align}
And at an observation at $t_k$ the filter equations are,
\begin{align}
& \hat{x}_{t_k}^{t_k^+} = \hat{x}_{t_k}^{t_k^-} + K_{t_k}(y_{t_k} - \hat{x}_{t_k}^{t_k^-}), \nonumber \\
& P_{t_k}^{t_k^+} = RP_{t_k}^{t_k^-}/(P_{t_k}^{t_k^-} + R), \nonumber \\
& K_{t_k} = P_{t_k}^{t_k^-}/(P_{t_k}^{t_k^-} + R).
\end{align}
The process noise covariance $Q$ can take one of three values \--- the best estimate $\hat{x}_{t_{k-1}}^{t_{k-1}+}$ using complete state equation $dx = (\alpha - x)dt$, the evolving estimate $\hat{x}_t^t$, or a constant value that has to be guessed. In the first two cases, the estimate converges to data (Fig. \ref{fig:illus_kf}a). However, when Q is taken as unity, the convergence is poor. (Fig. 1a). 
The mean square estimation error and steady state error covariance for different constant values of $Q$ as well as for the case where this is updated is shown in Fig. \ref{fig:illus_kf}b and Fig. \ref{fig:illus_kf}c respectively. When $Q$ is small, indicating confidence in the process model, the convergence is poor. When $Q$ is large, indicating less confidence in the process model, the convergence improves due to more reliance on the measurement. Choosing $Q$ from the Langevin model is a natural way to get good convergence. $\msquare$

For the system in Eqn. (\ref{CDsystem}), standard properties of stability apply to the Kalman Filter in Eqns. (\ref{predict}-\ref{Kalman gain})~\cite{jazwinski1970stochastic}. 
In particular, observability is a necessary condition for the stability.
The variation where the process noise covariance is set to the time-varying estimate rather than just its value before the interval should not qualitatively change the evolution of ($\hat{x}^t_t$, $P^t_t$).
This is because $P^t_t$ dynamics do not affect $\hat{x}^t_t$ in the interval between observations and the influence of $\hat{x}^t_t$ on $P^t_t$ is only as a time-dependent forcing term.
Finally, the proper working of the EKF is contingent on the closeness of the linearized approximation to the actual nonlinear system~\cite{speyer2008stochastic, anderson1979optimal}.

 \section{Parameter Estimation for Biomolecular Systems}
\label{sec3_kf}
Next, we investigate the performance and properties of this choice of process noise covariance through biomolecular circuit examples below.

\noindent \textit{Example 2. Signaling System} One of the widely occurring examples of biomolecular systems is of a signaling system in which a protein can exist in two different states ($A_0$ and $A_1$) and they interconvert among each other depending upon the forward and backward reaction rate constants~\cite{stock2000two}. 

We generate data using exact stochastic simulation of the Chemical Master Equation which captures the randomness in the process, both in terms of the occurrence of next reaction and the selection of the next reaction. This generates data with non-uniform discrete time samples. To mimic realistic conditions, measurement noise is added to the output as a uniform integer random variable in $[-5, 5]$, generated using \textit{randi} command in MATLAB. It is assumed that the statistics of the measurement noise
are known. Furthermore, since the number of molecules is always positive, any negative value generated is replaced by zero.

\begin{figure}[thpb]
      \centering
     \hspace*{-0.3cm} \includegraphics[scale=0.4]{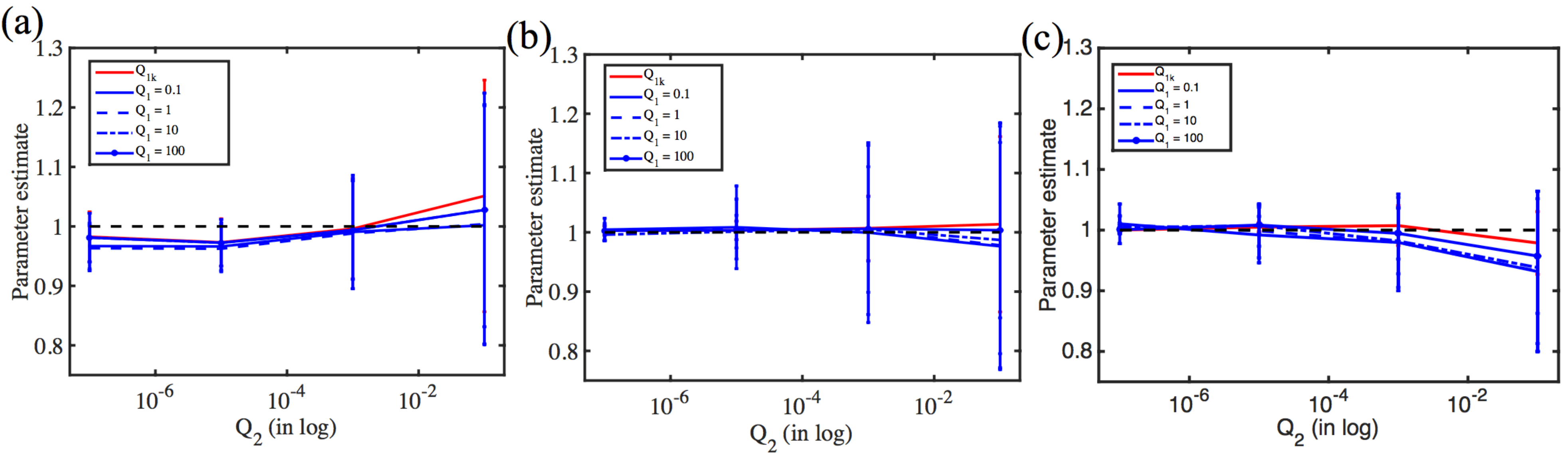}
      \caption{Effect of $Q_2$. Mean estimate of last 100 time points along with standard deviation for fixed $Q_1$, as well as updated $\hat{Q}_{1k}$ with an ensemble size $N = 10$ is plotted, the dashed black line shows the value of the unknown parameter used in simulation to generate data for a. $A_T = 10$, b. $A_T = 100$ and c. $A_T = 1000$.}
      \label{fig:q2}
   \end{figure}

The stochastic model of this system based on the Langevin equation is given by,
\begin{align}
& dA_1(t)=[K_{1}(A_{T} - A_1(t)) - K_{2} A_1(t)]dt + \sqrt{K_1(A_T - A_1(t))}d\beta_1(t) - \sqrt{K_2 A_1(t)}d\beta_2(t), \nonumber \\
& y_{k}=A_{1,k}+v_{k},
\end{align}
where $v_k$ is the measurement noise, $y_k$ is the measurement and $\beta_i(t)$ are independent zero mean unit variance brownian motion processes. 
The total number of molecules of this protein, existing in either state is constant, given by $A_T=(A_0+A_1)$. There are three parameters of this system $A_T$, $K_1$, and $K_2$. Extending these parameters as states and linearizing around estimates yields that the local observability matrix is full rank only if one parameter is unknown. For this reason, we choose that $A_T$ and $K_1$ is known and $K_2$ is to be estimated.
The process noise covariance matrix is of a diagonal structure following Langevin formalism, $Q = \begin{bmatrix}
Q_1 & 0\\
0 & Q_2
\end{bmatrix},$ where $Q_1$ is updated at each iteration using best estimates from complete nonlinear model as,
\begin{align}
\hat{Q}_{1,k}=K_{1}(A_{T} - \hat{A}^+_{1,k-1}) + \hat{K}^+_{2,k-1}\hat{A}^+_{1,k-1},
\label{Eqn:Qupdate_twostate}
\end{align}
and $Q_2$ is a small positive number ($10^{-7}$). In principle, if this unknown parameter were constant, corresponding process noise variance could be chosen as zero. However, as noted in~\cite{anderson1979optimal} and to allow for the possibility of time-varying parameters, a small value is chosen. Different values of this variance have been taken and their effect on parameter estimate is shown in Fig. \ref{fig:q2}. With higher values of this variance, the parameter estimates have higher variance and the mean parameter estimate shifts away from the true parameter value.

\begin{figure}[thpb]
      \centering
    \hspace*{-0.3cm}\includegraphics[scale=0.42]{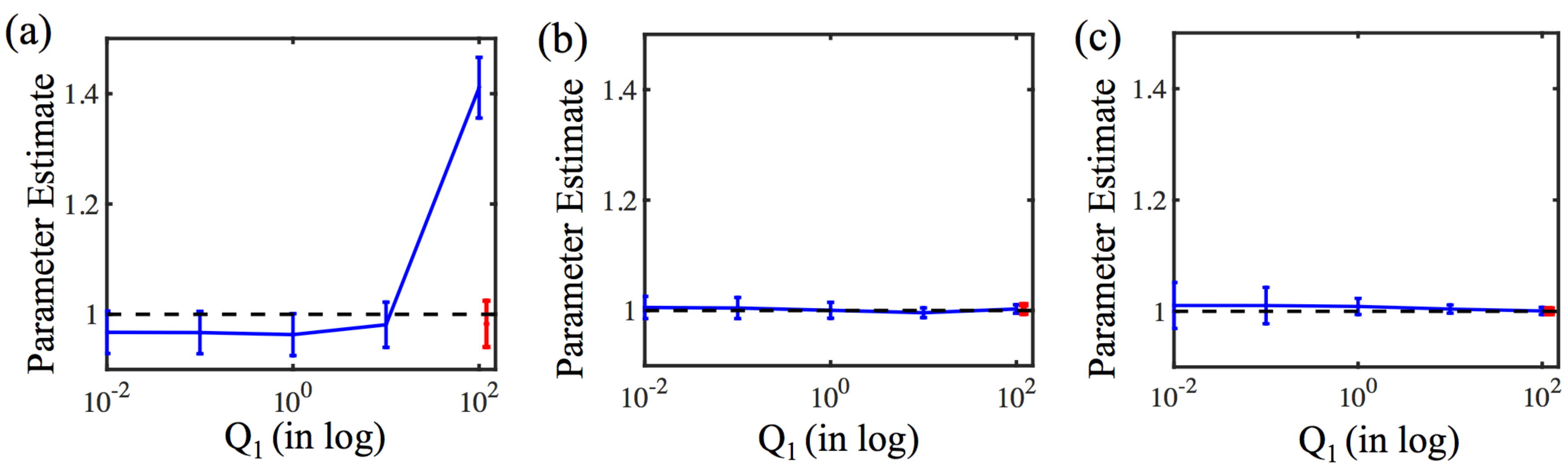}
      \caption{Estimated parameter $K_2$ for signaling system. Dashed black line shows true value of the parameter. Mean estimate of last 100 time points along with standard deviation for fixed $Q_1$ (blue line) as well as updated $\hat{Q}_{1k}$ (red dot) with an ensemble size $N = 10$ is plotted for a. $A_T = 10$, b. $A_T = 100$, and, c. $A_T = 1000$. $Q_2 = 10^{-7}$ in all three cases.}
      \label{fig:HEKF_1_kf}
   \end{figure}
We compared the parameter estimate for $K_2$ obtained with updated process noise $Q_{1k}$, with fixed value of process noise $Q_1$ (Fig. \ref{fig:HEKF_1_kf}) for three different values of $A_T$ = $10, 100, 1000$.
We find that the parameter estimate obtained from an update of the process noise $Q_1$ is reasonably close to the actual value in all three cases.
For the case of $A_T$ = $10$, the estimated value with a fixed $Q_1$ are closer to the actual estimate than that with an updated $Q_1$.  However, the choice of fixed value of $Q_1$ is unclear and our proposed method allows a good estimate without having to guess $Q_1$. $\msquare$

\noindent \textit{Example 3. Limit Cycle Oscillations}
Next, as a more complex example, we consider a biomolecular circuit exhibiting limit cycle oscillations (the repressilator~\cite{elowitz2000synthetic}).
This is a benchmark for biomolecular oscillator designs.
As above, the data are generated using the Gillespie algorithm for the associated Chemical Master Equation.\\
\vspace*{-1mm}
\indent The model based on the Langevin formalism is given by,
\vspace*{-1mm}
\begin{align}
& dm_i(t) = [\alpha_{0}+\frac{\alpha}{1+(p_j(t)/K)^n} - \gamma_m m_i(t)]dt + \sqrt{\alpha_{0}+\frac{\alpha}{1+(p_j(t)/K)^n}} d\beta_1(t) \nonumber \\
& - \sqrt{\gamma_m m_i(t)} d\beta_2(t),\nonumber \\
& dp_i(t) = [\beta m_{i}(t) - \gamma_{p}p_{i}(t)]dt + \sqrt{\beta m_{i}} d\beta_3(t) \nonumber \\
& - \sqrt{\gamma_{p}p_{i}(t)} d\beta_4(t), 
\end{align}
where $m_i$ and $p_i$ ($i = 1,2,3$) are the concentrations of mRNA and protein respectively. The $i^{th}$ mRNA is repressed by $j^{th}$ protein in a cyclic manner such as, for $i = 1$, $j = 3$, for $i=2$, $j=1$ and for $i=3$, $j=2$. We note that the functions used here such as $\alpha/(1+(p_j(t)/K)^n)$ are lumped from the underlying reaction schemes.
\begin{figure}[thpb]
      \centering
      \includegraphics[scale=0.42]{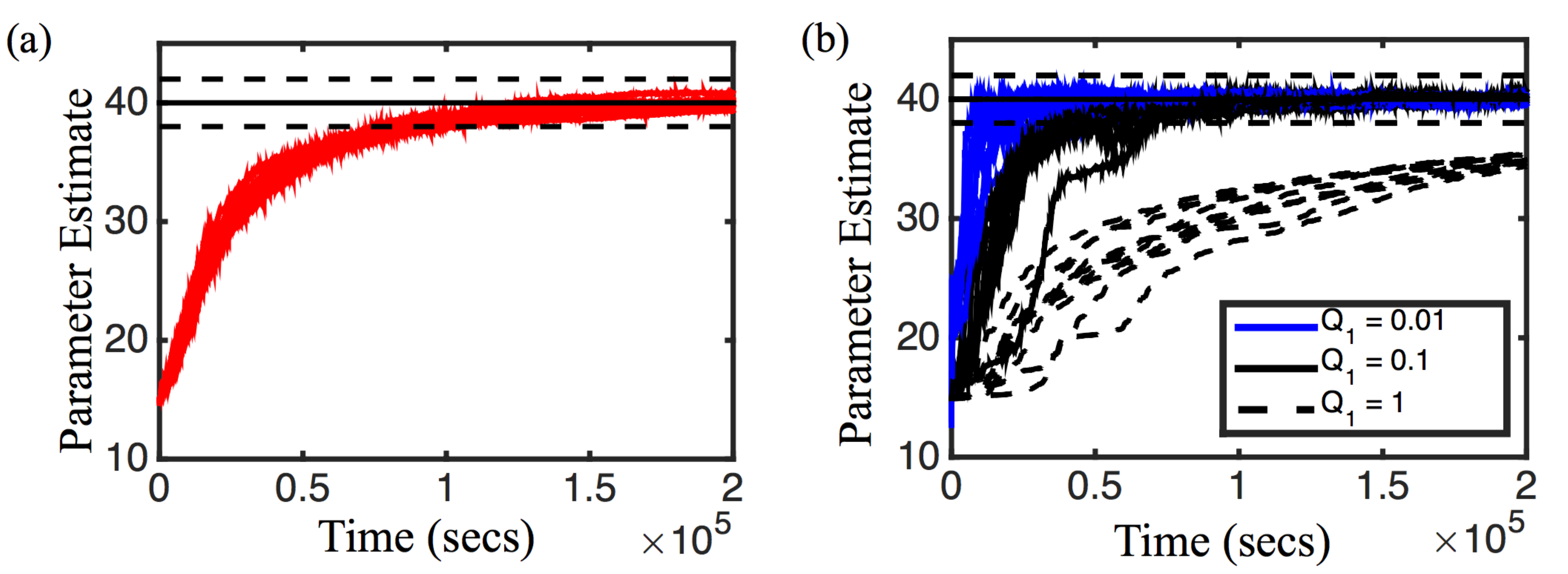}
      \caption{Estimation of parameter $K$ in repressilator model. Black solid and dashed horizontal lines represent true parameter value and $\pm$ $5 \%$ band of true parameter value. For each choice of process noise covariance, the filter algorithm is run 10 times. a. Parameter estimate using updated $Q_{i,k}$. b. Parameter estimate using fixed $Q$ of the structure $Q_1I_{6 \times 6}$. }
      \label{fig:repressilator_kf}
   \end{figure}
In principle, it should be possible to consider a model without this lumping, but it would be relatively larger.
For reasons of simplicity and because this is a model of such circuits that is in common use, we consider the above equations as a model.
We consider a linearization of the above equations and find that it is observable for the case when all protein concentrations ($p_i$) can be measured, which is realistic, and the only one parameter is unknown.
If more than one parameter is extended as state, it is unobservable. We choose parameter $K$ to be unknown, and other parameters known.\\
\indent The updated $Q$ based on Langevin formalism, $Q = \begin{bmatrix}
Q_{i,k} & 0\\
0 & q_2
\end{bmatrix}$, where $Q_{i,k} \in \mathcal{R}^{6 \times 6}$ is a block diagonal matrix with diagonal entries being state dependent,
\begin{align}
\hat{Q}_{1,k} ={}& \gamma_{m} \hat{m}^{+}_{1,k-1}+\alpha_{0}+ \alpha/(1+(p_{3}/\hat{K}^{+}_{k-1})^n), \nonumber \\
\hat{Q}_{2,k} ={}& \gamma_{p}\hat{p}^{+}_{1,k-1}+\beta \hat{m}^{+}_{1,k-1},
\label{Eqn:Qupdate_repress}
\end{align}
and similar for other equations and $q_2$ is a small positive number ($10^{-7}$) to allow for the possibility of time-varying parameters.

We compared the estimate of $K$ using updated $Q$ (Eqn. \ref{Eqn:Qupdate_repress}) with different fixed $Q$ (Fig. \ref{fig:repressilator_kf}). We find that, although the updated $Q$ does not provide fastest convergence, still it gives good estimates and has advantage that $Q$ does not have to be chosen by trial and error. $\msquare$

\noindent \textit{Example 4. Heat Shock Response of E. coli} 

Next, we take an example of the heat shock response model of \textit{E. coli}. 
While exposed to higher temperatures \textit{E. coli} shows a different response than normal temperature range ($30^o - 37^o$C) due to activation of heat shock genes. This is simulated using the Euler-Maruyama method with a time step of $0.001$ for $150000$ time instants. 
Every $100$th point is taken and a measurement noise corresponding to a zero mean unit variance white Gaussian noise is added to it to obtain the data.The stochastic model of this based on Langevin formalism is given by, 

\begin{align}
& dD_t = \Big[ K_d\dfrac{S_t}{1+\frac{K_sD_t}{1+K_uU_f}} - \alpha_dD_t \Big]dt + \sqrt{K_d\dfrac{S_t}{1+\frac{K_sD_t}{1+K_uU_f}}}d\beta_1(t) - \sqrt{\alpha_dD_t}d\beta_2(t), \nonumber \\
& dS_t = \Big[\eta(T) - \alpha_0S_t - \alpha_s \dfrac{\frac{K_sD_t}{1+K_uU_f}}{1 + \frac{K_sD_t}{1+K_uU_f}}S_t \Big]dt + \sqrt{\eta(T)}d\beta_3(t) - \sqrt{\alpha_0S_t}d\beta_4(t) \nonumber \\
& - \sqrt{\alpha_s \dfrac{\frac{K_sD_t}{1+K_uU_f}}{1 + \frac{K_sD_t}{1+K_uU_f}}S_t}d\beta_5(t), \nonumber \\
& dU_f = \Big[ K_T(P_t - U_f) - (K(T) + K_{fold})D_t \Big]dt + \sqrt{K_TP_t}d\beta_6(t) - \sqrt{K_TU_f}d\beta_7(t) \nonumber \\
& - \sqrt{(K(T) + K_{fold})D_t}d\beta_8(t).
\end{align} 
\begin{figure}
      \centering
     \hspace*{-0.3cm} \includegraphics[scale=0.52]{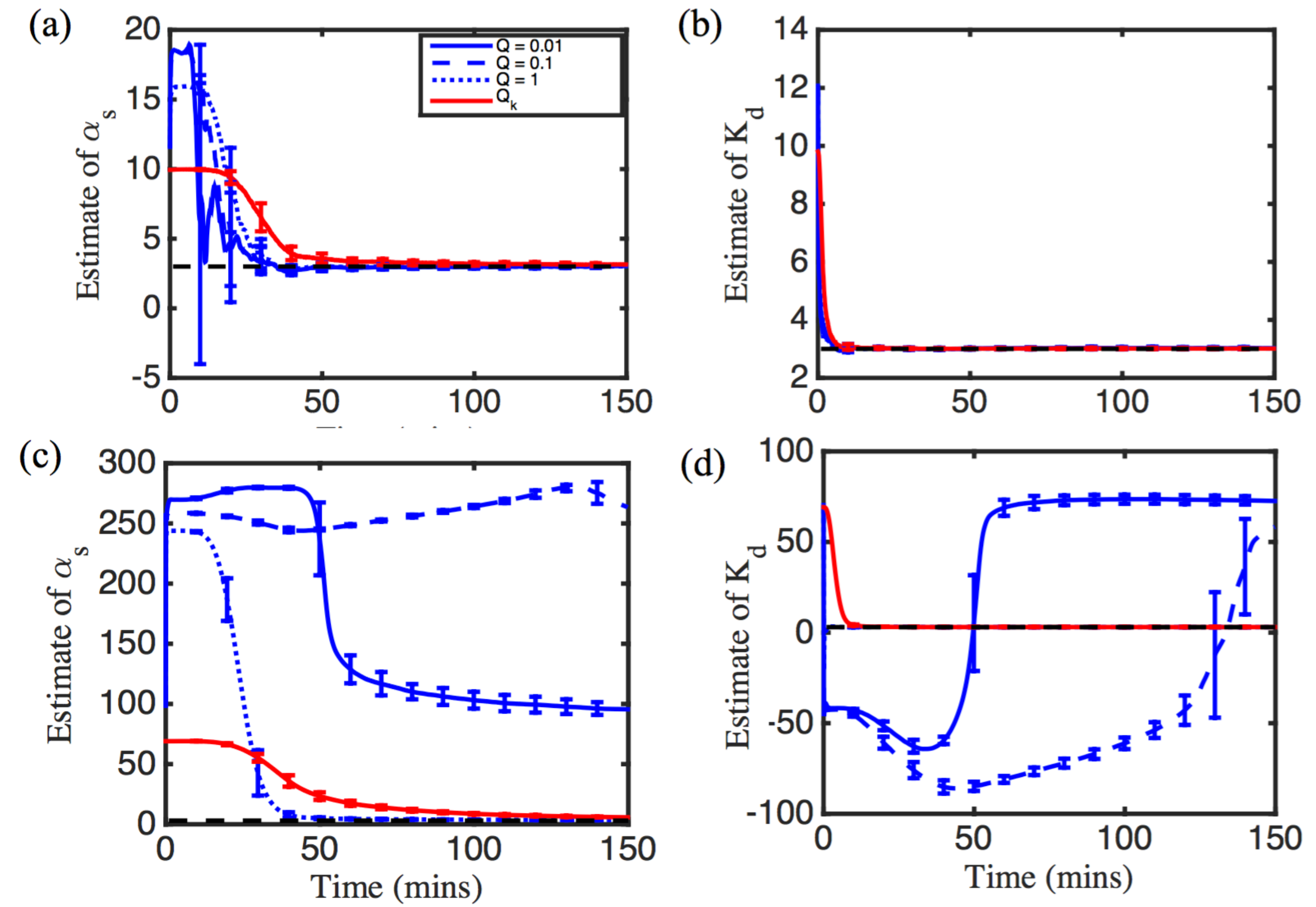}
      \caption{Estimated parameter $\alpha_s$ and $K_d$ of \textit{E. coli} heat shock response model. Blue lines represent fixed process noise covariance, red line represents updated process noise covariance and dashed black line represents true value of the parameter. a. Estimate of $\alpha_s$, b. estimate of $K_d$ when initial guess of parameters are set as $10$. c. Estimate of $\alpha_s$, d. Estimate of $K_d$ when initial guess of parameters are set as $70$. }
      \label{fig:heatshock}
   \end{figure}
This is derived from a reduced order model proposed by~\cite{el2006advanced}. We assume that measurements of the variables $D_t$ and $S_t$ are available. The linearized model with parameter extended as states show that the system is observable while two of the parameters are unknown. We choose $\alpha_s$ and $K_d$ to be unknown and rest of the parameters known. 

The updated process noise covariance matrix, $Q = \left[
\begin{array}{c|c}
Q_{i,k} & 0 \\
\hline
0 & \eta
\end{array}
\right]$, where the diagonal entries of $Q_{i,k}$ are given by,
\begin{align}
& Q_{1,k} = K_{d,k-1}^+\dfrac{S_{t,k-1}^+}{1+\frac{K_sD_{t,k-1}^+}{1+K_uU_{f,k-1}^+}} + \alpha_dD_{t,k-1}^+, \nonumber \\
& Q_{2,k} = \eta(T) + \alpha_0S_{t,k-1}^+ + \alpha_{s,k-1}^+ \dfrac{\frac{K_sD_{t,k-1}^+}{1+K_uU_{f,k-1}^+}}{1 + \frac{K_sD_{t,k-1}^+}{1+K_uU_{f,k-1}^+}}S_{t,k-1}^+, \nonumber \\
& Q_{3,k} = K_TP_t + K_TU_{f,k-1}^+ + (K(T) + K_{fold})D_{t,k-1}^+t. \nonumber
\end{align}
And $\eta$ is a positive diagonal matrix $Q_2I_{2 \times 2}$, where $Q_2 = 10^{-7}$ to allow time-varying parameters. In case of a fixed process noise covariance matrix, $Q_{i,k}$ is replaced by $QI_{3 \times 3}$, where $Q$ is a constant.

We compared the parameter estimates obtained with fixed and updated process noise covariance choices (Fig. \ref{fig:heatshock}). We observe while both choices lead to satisfactory parameter estimates with the initial guess for parameter is closer to true value (Fig. \ref{fig:heatshock}a,b), the parameter estimates do not converge to true value in case the initial guess is far from the true value choosing fixed process noise covariance (Fig. \ref{fig:heatshock}c,d). However, with updated $Q$ the parameter estimates converge to true value even if the initial guess is far off.

\noindent \textit{Example 5. Experimental Data of Negative Transcriptional Feedback Circuit}

\begin{figure}[h]
\centering
\hspace*{-0.4cm}\includegraphics[scale=0.5]{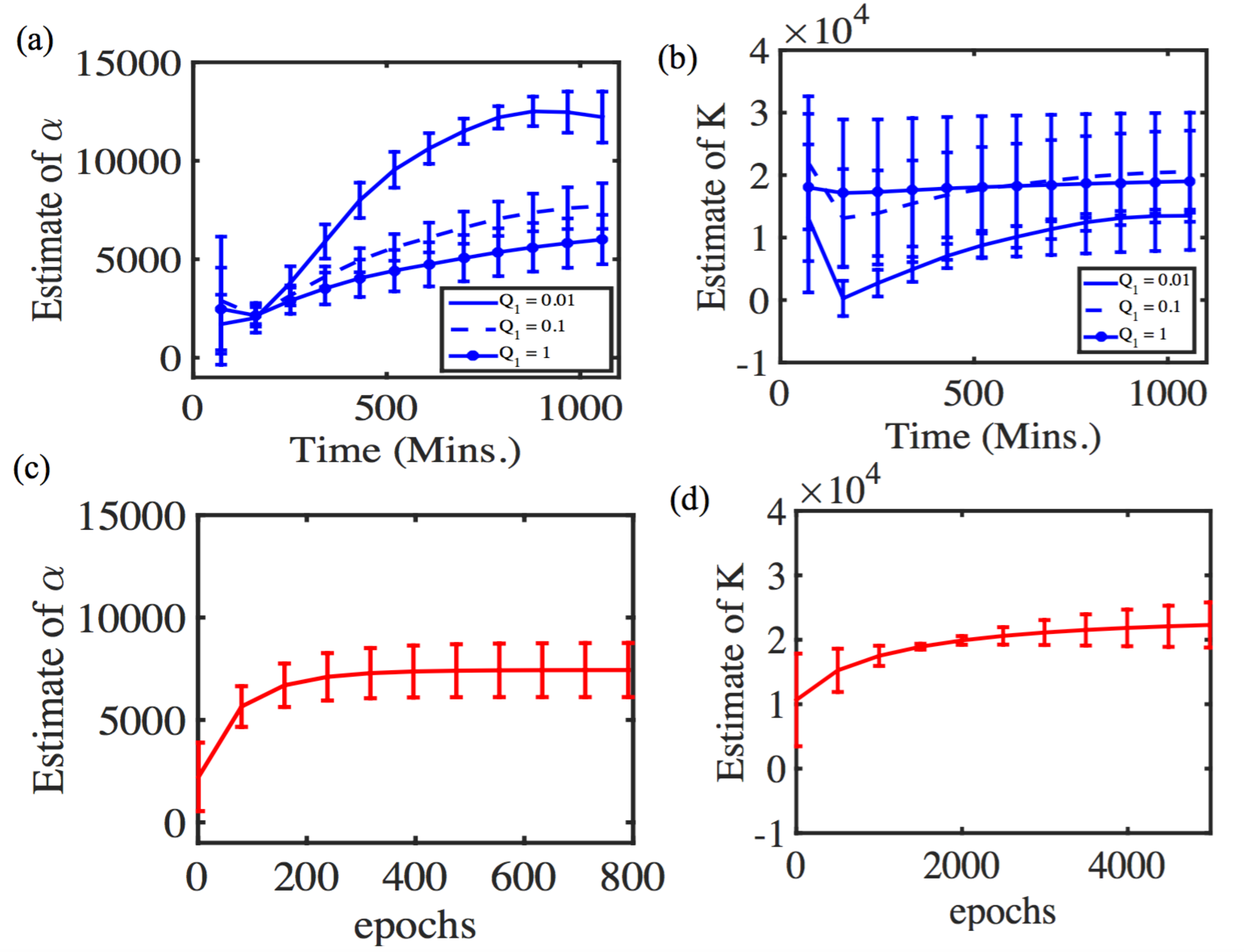}
\caption{Estimation of parameters from experimental data of negative transcriptional feedback circuit. Blue lines represent estimate using different fixed values of process noise covariances for, a. $\alpha$  and b. $K$. Red lines represent estimate using updated process noise covariance for, c. $\alpha$ and d. $K$. The value of measurement noise covariance, $R$ is taken as $0.1$ in all the cases.}
\label{fig:negfeed_kf}
\end{figure}

Finally, we consider an example of a negative transcriptional feedback circuit where a protein can repress its own production. This kind of network motif occurs in over $40\%$ of transcription factors in \textit{E. coli}~\cite{rosenfeld2002negative}. We acquired experimental data of this network from~\cite{sen2013temperature}. These data are available at a sampling interval of $10$ minutes.\\
\indent The model based on Langevin formalism is given by,
\begin{align}
& dx(t) = [\frac{\alpha}{1+x(t)/K} - \gamma x(t)]dt + \sqrt{\frac{\alpha}{1+x(t)/K}}d\beta_1(t) - \sqrt{\gamma x(t)}d\beta_2(t), \nonumber \\
& y_k = x_k + v_k,
\end{align}
where, $y_k$ are measurements, $\beta_i(t)$ and $v_k$ are the process and measurement noise respectively. In this model also nonlinear terms are lumped together, but for simplicity we consider this to be the process model. The measurement sampling interval is chosen to be as small as possible but limited by experimental parameters such as the bleaching of the fluorescent reporter. Here, the parameters $\alpha$, $\gamma$, and $K$ are to be estimated. But, with three unknown parameters the extended linearized system becomes unobservable and only one unknown  parameter can be estimated. To tackle this, $\gamma$ is computed from the maximum value of forward difference in growth rate data which are smoothened using a moving average filter (\textit{smooth} function in MATLAB with default options). Now, the remaining parameters $\alpha$ and $K$ are estimated in an iterative method. First, $K$ is assumed to be known and $\alpha$ is estimated. Then, using the final estimate of $\alpha$ as known parameter, $K$ is estimated. This process is repeated until estimated parameters in each iteration are close to the estimates in previous iteration. To overcome the problem of slow convergence in case of process noise covariance updated based on the estimates, Algorithm \ref{algo2} is used~\cite{meskin2013parameter}.
\begin{algorithm}[H]
\caption{Algorithm for estimating parameters for negative feedback model with process noise covariance update}\label{algo2}
\begin{algorithmic}[1]
\STATE Set iteration counter, $i = 1$
\STATE Set initial guess of parameter $K$ assuming it is known
\STATE Set epoch counter, $j = 1$
\STATE Apply Algorithm 1 (in manuscript) to data set $y_k$ $(k = 1, 2, ... ,N)$ and calculate last estimate of $\alpha$
\STATE If $j > j_{max}$ then stop, otherwise re-initialize Algorithm 1 with $\hat{\alpha}(0) = \hat{\alpha}(N)$
\STATE Calculate last estimate of $\alpha$ and assuming that as known parameter, repeat steps 3 to 5 to estimate $K$
\STATE Repeat steps 2 to 6 until $|K_i - K_{i-1}|$ and $|\alpha_i - \alpha_{i-1}|$ are sufficiently small or $i = i_{max}$
\end{algorithmic}
\end{algorithm}
\indent We compared the estimated parameters using both updated and fixed choices of $Q$ in Fig. \ref{fig:negfeed_kf}. We obtained similar estimates using updated $Q$ and using $Q_1 = 0.1$ and $Q_1 = 1$, but the estimates using $Q_1 = 0.01$ are different. These simulations point towards the usefulness of using the information in process noise for estimation. $\msquare$
\section{Performance Of Filter}
\label{section4}
An important part of filter design is to check its performance. Here, we check the performance of the filter using the following two measures.\\ 
\textit{A. Whiteness of Innovation Sequence}
Checking the statistics of the innovation sequence is often used as a metric to test for the optimality of the filter, both in the standard and extended versions~\cite{anderson1979optimal}. For an optimal filter, the innovation sequence is a Gaussian white noise~\cite{kailath1968innovations}. The innovation sequence ($\nu_k$) is given by, $\nu_k = y_k - h(\hat{x}_k)$.
We estimated normalized autocorrelation of innovation sequences with time-shift $k=1,2,...,100$ using,
\begin{align}
\hat{\rho}_k = \frac{\hat{C}_k}{\hat{C}_0}, \hat{C}_k=\dfrac{1}{N}\sum_{i=k}^{N}\nu_i\nu_{i-k}^T,
\label{normalized autocorrelation}
\end{align}
where $N$ is the number of sample points~\cite{mehra1970identification}. $\hat{\rho}_k$ being normally distributed, the $95\%$ confidence limits are $\pm \dfrac{1.96}{\sqrt{N}}$. The filter is more optimal if less $\hat{\rho}_k$ values lie outside these confidence limits.\\

\begin{figure}[thpb]
      \centering
     \hspace*{-0.3cm} \includegraphics[scale=0.45]{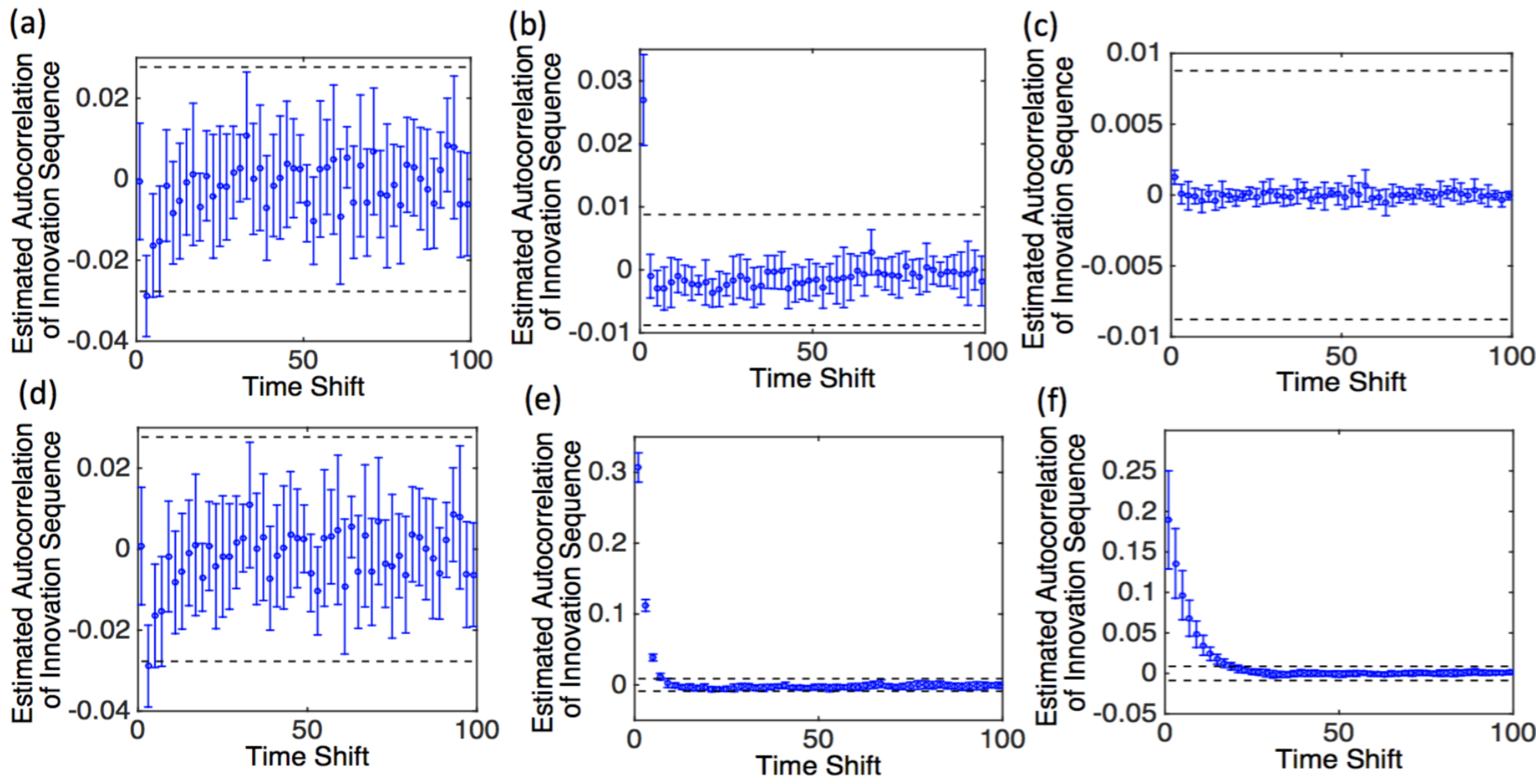}
      \caption{Whiteness of the innovation sequence for Example 2. The blue lines are mean estimated autocorrelation along with standard deviation and the black lines denote $95\%$ confidence limit for subfigure a to f. Estimated autocorrelation of innovation sequence for different values of time-shift, a.  for $A_T = 10$, using updated $Q_{1k}$.  b. for $A_T = 100$, using updated $Q_{1k}$. c. for $A_T = 1000$, using updated $Q_{1k}$. d. for $A_T = 10$ using fixed $Q_1=1$. e. for $A_T = 100$ using fixed $Q_1=1$. f. for $A_T = 1000$ using fixed $Q_1=1$.}
      \label{fig:optimality_1}
   \end{figure}

\begin{table}[thpb]
\caption{Percentage of estimated autocorrelation values outside $95\%$ confidence limits for biomolecular systems.}
\centering
\begin{tabularx}{\textwidth}{|X|X|X|X|X|}
 \hline
 Example 2 & $Q_k$ & $Q_1=1$ & $Q_1=10$ & $Q_1=100$\\
 \hline
 $A_T=10$ & 22\% & 45\% & 38\% & 42\% \\
 $A_T=100$ &  0\% & 100\% & 40\% & 0\%\\
 $A_T=1000$ & 0\% & 100\% & 100\% & 39\%\\
  \hline
  \multirow{ 2}{*}{Example 3} & $Q_k$ & $Q_1=0.01$ & $Q_1=0.1$ & $Q_1=1$\\
   \cline{2-5}
   & 12\% & 100\% & 100\% & 31\% \\
   \hline
    \multirow{ 2}{*}{Example 4} & $Q_k$ & $Q_1=0.01$ & $Q_1=0.1$ & $Q_1=1$\\
    \cline{2-5}
    & 38\% & 100\% & 87\% & 60\% \\
    \hline
   \multirow{ 2}{*}{Example 5} & $Q_k$ & $Q_1=0.01$ & $Q_1=0.1$ & $Q_1=1$\\
    \cline{2-5}
    & 19\% & 73\% & 72\% & 54\% \\
    \hline
\end{tabularx}
\label{table_1}
\end{table}
\indent We performed this test on the innovation sequence obtained from the above three examples for different fixed as well as updated process noise covariances (Table \ref{table_1}). The corresponding trajectories are shown in Fig. \ref{fig:optimality_1} for Example 2. We find that the filter is closer to optimality condition with the updated process noise covariance in all the cases.

\begin{figure}[thpb]
\centering
\hspace*{-0.4cm}\includegraphics[scale=0.5]{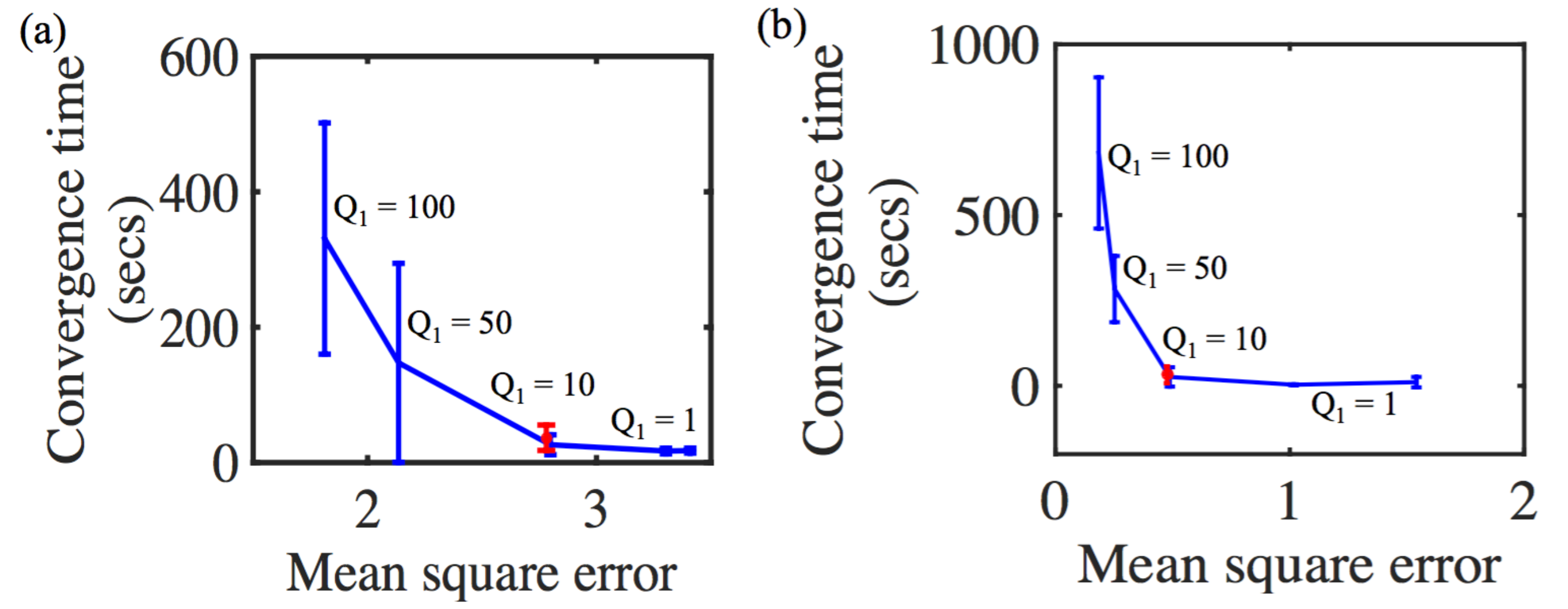}
\caption{Trade-off between convergence time and MSE for Example -2, $A_T =10$ for fixed $Q_1$ (blue line) as well as updated $\hat{Q}_{1k}$ (red dot). Convergence time is considered to be the time taken to reach $\pm 5\%$ band of true parameter value. Measurement noise ($v_k$) is an uniform integer random variable a. in $[-5, 5]$. b. in $[-1, 1]$.}
\label{fig:tradeoff_kf}
\end{figure}

\noindent \textit{B. Mean Square Estimation Error}
The mean square error (MSE) in estimation is another important criterion in performance of a filter,
\begin{equation}
MSE = \sqrt{\frac{1}{N}\sum_{k=0}^{N}(y_k - h(\hat{x}_k))^2},
\end{equation}
\begin{table}[hpt]
\caption{Mean square estimation errors for biomolecular systems.}
\centering
\begin{tabularx}{\textwidth}{|X|X|X|X|X|}
\hline
 Example 2 & $Q_k$ & $Q_1=1$ & $Q_1=10$ & $Q_1=100$\\
 \hline
 $A_T=10$ & 2.85 & 3.37 & 2.87 & 1.87 \\
 $A_T=100$ &  2.78 & 4.75 & 3.66 & 2.78\\
 $A_T=1000$ & 2.76 & 7.39 & 4.92 & 3.64\\
  \hline
  \multirow{ 2}{*}{Example 3} & $Q_k$ & $Q_1=0.01$ & $Q_1=0.1$ & $Q_1=1$\\
   \cline{2-5}
 & 2.98& 4.21& 3.55& 3.13 \\
  \hline
  \multirow{ 2}{*}{Example 4} & $Q_k$ & $Q_1=0.01$ & $Q_1=0.1$ & $Q_1=1$\\
    \cline{2-5}
    & $0.18$ & $12.51$ & $7.96$ & $5.35$ \\
    \hline
   \multirow{ 2}{*}{Example 5} & $Q_k$ & $Q_1=0.01$ & $Q_1=0.1$ & $Q_1=1$\\
    \cline{2-5}
    & $4.1 \times 10^{-3}$ & $6.4 \times 10^3$ & $990.7$ & $227$ \\
    \hline
\end{tabularx}
 \label{table_2}
\end{table}%
where $h$ is output function of states. We compared the MSE obtained for above examples using both updated and fixed process noise covariances in Table \ref{table_2}. For Examples  2 and 3 the added measurement noise ($v_k$) was a uniform integer random variable in $[-5, 5]$, and its standard deviation can be computed as, $\sqrt{E(v^2) - (E(v))^2} = 3.16$ which is close to the MSE using updated $Q$ (Table \ref{table_2}). For Example 4 the MSE with updated $Q$ is much lower than fixed choices.\\
\indent We also investigated a possible trade-off between MSE and parameter convergence time for different values of process noise covariances~\cite{speyer2008stochastic}. We further investigated this for Example 2 with total concentration, $A_T = 10$ and different measurement noise levels (Fig. \ref{fig:tradeoff_kf}). We note that the data point corresponding to the updated process noise covariance lies near the vertices of these hyperbolic curves.

\section{Discussion}
We have adapted a Hybrid Extended Kalman Filtering technique for biomolecular systems where the process noise covariance, as in the Langevin formalism, depends on state and/or parameters.
We emphasize the usefulness of this information and update the covariance of the process noise with each observation.
We applied this method to estimate parameters in simple models of biomolecular circuits, including those that display limit cycle oscillations, using data generated from Monte Carlo stochastic simulations, Euler-Maruyama simulations as well as to experimental data of a negative transcriptional feedback circuit.
We find that innovation sequence in this framework can be white, indicating that the filter is closer to optimality condition, in contrast to setting a fixed, though unknown, value of process noise covariance.
This updated process noise covariance can also provide a balance between the mean square estimation error and parameter convergence time.
This framework naturally provides a choice for the process noise covariance, which is an important part of Kalman Filter design.

We note an interesting tradeoff between mean square error and parameter convergence time for different choices of fixed noise covariances for Example 2 (Fig. \ref{fig:tradeoff_kf}).
For higher noise covariances, the mean square error reduces, but parameter convergence is slow.
The opposite happens for lower noise covariances.
The choice of noise covariance updated based on estimates can provide a balance.

\begin{figure}[thpb]
      \centering
     \hspace*{-0.3cm} \includegraphics[scale=0.5]{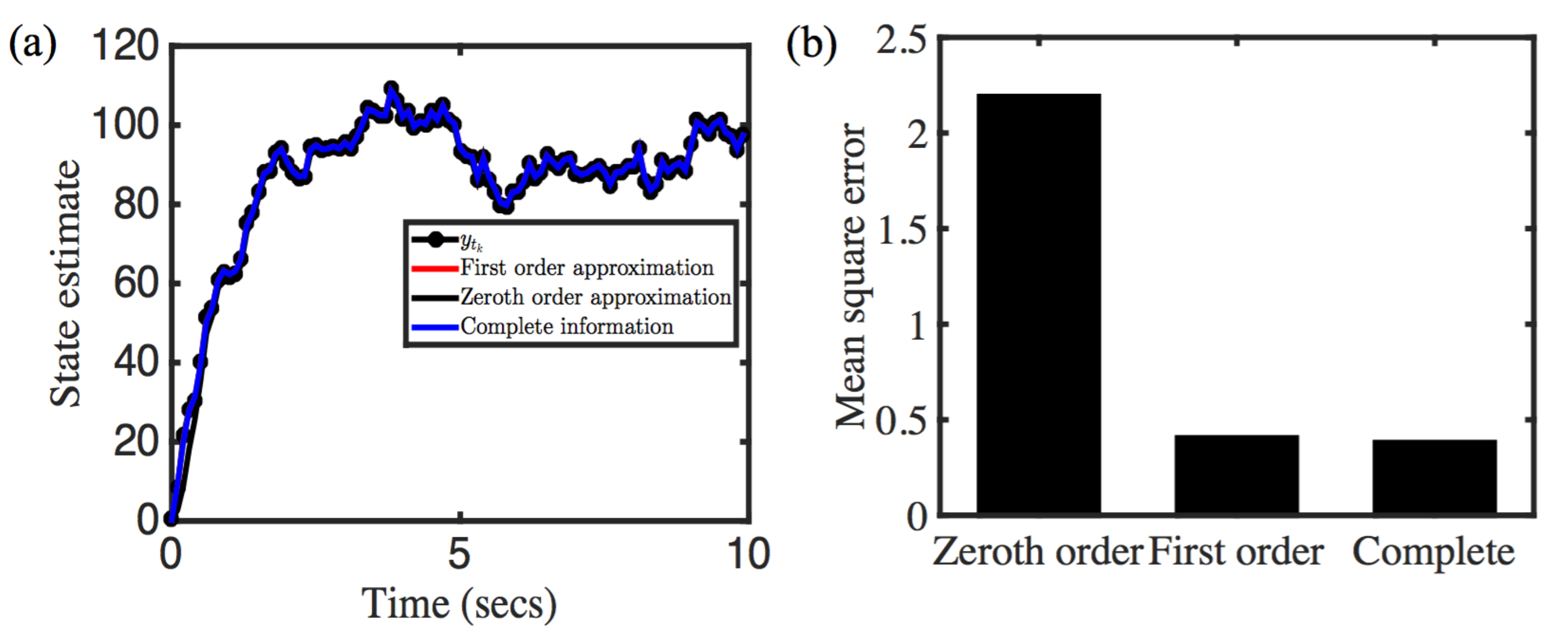}
      \caption{Comparison of three cases of updated process noise covariance for Example 1. a. state estimates, b. corresponding mean square errors in estimation.}
      \label{fig:approxQ}
   \end{figure}

We note that the process noise may be approximated in different ways through the use of higher order approximations.
The cases where it is approximated by the zeroth order term and the case where the complete information is added has been noted in Section \ref{sec2}.
Another approximation is to consider upto first order~\cite{jimenez2002linear}.
In this case, the filter equations for ($\hat{x}^t_t$, $P^t_t$) change.
For Example 1, these become,
\begin{align}
& d\hat{x}^t_t = (100-x)dt,\nonumber \\
& dP^t_t = \Big[(-2 + \frac{1}{4\hat{x}^t_t(0)})P^t_t + \frac{(\hat{x}^t_t + \hat{x}^t_t(0))^2}{4\hat{x}^t_t(0)} \Big]dt. 
\end{align}
Performance of this filter is similar to others (Fig. \ref{fig:approxQ}).
Care needs to be taken in transients, where due to terms like $\dfrac{1}{\hat{x}^t_t(0)}$, large values may arise if $\hat{x}^t_t(0)$ is small.

An open issue in this framework is the choice of noise covariance to allow time-varying parameters that has been investigated in case of Example 2. Further how this choice affects the mean square estimation error and convergence speed in general is to be examined. An important task for the future is to generalize the use of information of the dependence of process noise covariance on the state and/or parameters to other filtering techniques.
The filters discussed so far are approximate in the sense that the system model is linearized, to various degrees, before the linear filter is applied.
A possible approach is to develop the complete nonlinear filter and then approximate this nonlinear filter using truncation techniques~\cite{jazwinski1970stochastic}. 
Another approach is to consider Particle Filters or Cubature Kalman Filters, which have been reported for such continuous-discrete systems~\cite{xia2013new, arasaratnam2010cubature}. 
Particle Filters have also been used in biomolecular contexts~\cite{lillacci2012distribution, liu2012state}, with better performance than EKFs when multiple parameters are sought to be estimated.
Adding this information of the state/parameter-dependent process noise may further improve their performance.

\textit{A priori} knowledge of the system model including noise characteristics has been noted to be important in the design of Kalman Filters.
Here, we have added to this by emphasizing the use of the dependence of the process noise in biomolecular systems on parameters and/or states.
This dependence is obtained from the Chemical Langevin formalism of biomolecular system models.
We find that including this information can improve filter performance.
This framework should help in parameter estimation in biomolecular contexts as well as in other contexts where process noise is known to depend on states and parameters.

\section*{ACKNOWLEDGMENT}

We thank Prof. I. N. Kar for his valuable inputs. Research supported partially by Science and Engineering Research
Board grant SB/FTP/ETA-0152/2013.

\bibliographystyle{unsrt}
\bibliography{ref}
\end{document}